# Physical Processes in Compensated Negative-Ion Beams and Related Beam Transport Problems

I.A.Soloshenko[2], V.N.Gorshkov[1,2], and A.M.Zavalov[2]

[1]CNLS and Theoretical Division, Los Alamos National Laboratory, Los Alamos, NM, 87544
[2]Institute of Physics, National Academy of Sciences, Ukraine, Kiev

**Abstract**

This paper reviews resent theoretical and experimental research on physical processes in compensated negative-ion beams. A previous review was published by one of the authors in Ref. [1]. The present paper summarizes the new results of numerical calculations of the stationary electric field in negative-ion beams, and the resent results of experimental studies of suppression of ion-ion instability at low gas pressures.

As in Ref. [1], the review addresses the following three main issues of the physics of compensated beams:
1) Spatial charge compensation in steady beams;
2) Dynamic decompensation of spatial charge, which occurs as a result of beam current modulation due to plasma instabilities in the ion source;
3) Collective processes.

The results presented in this review enable one to determine how non-compensated spatial charge, nonresonant oscillations of the beam current, and collective oscillations of the ion-beam plasma influence the propagation properties of the beam. With this information, one can determine the optimal ion transportation parameters.

**Introduction**

Intense ion beams have large spatial charges. Consequently, ion transport is possible only if their charges are compensated by charges of the opposite sign. For negative ion beams, the spatial charge compensation can be achieved using positive ions that are produced by ionization of a neutral gas. In the process of the ionization, the beam becomes a two- or three-component quasi-neutral medium that is commonly called an ion-beam plasma. For a given concentration, $n_0$, of neutral gas particles, an ionization cross-section, $\sigma_i$, and an ion velocity, $v_-$, the time of ionization, $\tau$, can be estimated to be $\tau = 1/n_0 \sigma_i v_-$. The introduction of the compensating component mainly solves the problem of the spatial charge neutralization and, consequently, the problem of the beam transport. However, while the transport properties of a single-component beam are easily calculated using numerical techniques, the analysis of the multi-component system is rather complicated. Thorough experiments and numerical simulations are required to investigate the transport properties of multi-component ion beams. This research has been systematically performed at the Institute of Physics of National Academy of Sciences of the Ukraine for many years. For convenience, we have subdivided the results obtained into the following three parts:

1) The spatial charge compensation of a steady ion beam;
2) The dynamic decompensation of the beam occurring as a result of current modulation by nonresonant oscillations;
3) The collective processes in an ion-beam plasma.

**I. Spatial charge compensation in a steady ion beam**

The main difference between intense beams of negative ions and intense beams of positive ions is that negative ion compensation is performed by heavy positive ions. Consequently, the spatial charge of the negative ion beam may be overcompensated. (For positive ion beams overcompensation is possible only in the presence of external magnetic field.) This phenomenon of the spatial charge overcompensation is observed at gas pressure, $p$, that exceeds a critical value, $p_{cr}$. The related critical concentration of neutral particles, and, consequently, the critical pressure, can be easily determined from the balance equations written at this pressure for electrons and plasma ions:

$$n_-v_-n_{0cr}\sigma_i\pi r_0^2 = n_i v_i 2\pi r_0$$
$$n_-v_-n_{0cr}\sigma_e\pi r_0^2 = n_e v_e 2\pi r_0 \qquad (1)$$

where $n_-$, $n_i$, and $n_e$ - are the concentrations of the beam ions, the plasma ions, and the plasma electrons; $\sigma_e$ and $\sigma_i$ - are the generation cross sections of the electrons (including both the ionization and recharging processes) and the positive ions; $v_i$ and $v_e$ are the mean velocities of the ions and the electrons; $r_0$ is the beam radius. Taking into consideration the equation of quasi-neutrality

$$n_e + n_- = n_i, \qquad (2)$$

which is satisfied strictly enough, and the equations (1), one can easily obtain the following expression

$$n_{0\,cr} = \frac{2\,v_i}{v_-\sigma_i r_0\left(1 - \frac{v_i}{v_e}\frac{\sigma_e}{\sigma_i}\right)} \approx \frac{2\,v_i}{v_-r_0\sigma_i} \qquad (3)$$

If the gas pressure differs from the critical value, $p \neq p_{cr}$, two regimes should be considered.

1. For pressures lower than the critical value, $p < p_{cr}$, the system can be considered as having two-component, because the electric field pushes the electrons out of the beam, and the steady concentration $n_e(p)$ is less than $n_-$ by several orders of magnitude, $n_e(p) \ll n_-$, the concentration of positive ions, $n_i(p)$, being lightly less than $n_-$.
2. When the gas pressure exceeds the critical value, $p > p_{cr}$, the system has three components because the electric field holds slow electrons inside the beam. At high pressures, the concentration $n_e(p > p_{cr})$ may be greater than the beam concentration.

For optimal negative ion transport over noticeable distances, one must evaluate the electric field within the beam for both of these regimes. Let us first consider the regime of high pressure. In Ref. [2] the radial voltage drop, $\Delta\varphi$, (which is unambiguously related to the average radial electric field) was determined using the equation of electron balance

$$n_- v_- n_{0cr} \sigma_e \pi r_0^2 = 2\pi r_0 n_e v_e e^{-\frac{e\Delta\varphi}{kT_e}} \qquad (4)$$

where $T_e$ is the plasma electron temperature.

Usually, this equation rather adequately describes the real system because the lifetime of the free electrons significantly exceeds the time of the electron-electron collisions, and the distribution function of the electrons has sufficient time to reach equilibrium (the Boltzman's distribution). However, from this equation one can obtain only the relation between $\Delta\varphi$ and $kT_e$. It is not a solution of the problem because both $kT_e$, and $\Delta\varphi$ are undefined parameters. In Ref. [3] an equation of the energy balance for compensating electrons is proposed to use for determination of the value $\Delta\varphi$. It is assumed that the compensating electrons are slow electrons trapped in potential well of the beam, whereas electrons whose energy is sufficient to overcome the potential barrier, $e\Delta\varphi$, rush out of the beam, and their contribution to the spatial charge can be neglected.

The authors of Ref. [3] focused their attention on the following phenomenon. On the one hand, compensating electrons, which experience multiple oscillations in potential well, cannot leave the beam because of insufficient energies. On the other hand, they have to leave the beam in the course of time to prevent a permanent accumulation of the slow electrons otherwise the stationary state of the system should not be realized. Thus, some energy source heats up the trapped electrons (in the absence of such a source, the well depth would equal to zero). The beam itself provides this energy source. Coulomb collisions or different collective processes may cause the energy transfer from the beam ions to the electrons. To find a minimal value for $\Delta\varphi$ one should consider only the Coulomb collisions.

Let us estimate the energy per centimeter per second, $\Delta E_i$, that ions must transfer to the trapped electrons to enable all electrons to leave the beam. If $f(\varepsilon)$ is the energy distribution function of electrons created by the ion-induced ionization of the neutral, then at least

$$\Delta E_i \geq \int_0^{r_0} 2\pi r dr \int_0^{e\varphi} f(\varepsilon)(e\varphi - \varepsilon) d\varepsilon . \qquad (5)$$

When being heated up, each trapped electron accumulates energy rather gradually through the long-rang electron-ions forces acting within the Debye sphere, in opposite to neutral molecules, for example, energies of which can change step-wise in time due to the short-rang forces acting by collisions of the molecules with each other. As soon as the kinetic energy of an electron reaches the well depth, the electron will leave the beam quickly, and so its velocity vanishes out of the beam. This circumstance enables to write the expression (4) as the equality

$$\Delta E_i = \frac{\pi n_- n_e e^4 L r_0^2}{m v_-} \approx \int_0^{r_0} 2\pi r dr \int_0^{e\varphi} (e\varphi - \varepsilon) f(\varepsilon) d\varepsilon , \qquad (6)$$

where $L = 4\pi\Lambda$ and $\Lambda$ is Coulomb logarithm.
Using simplifying assumptions, $\varphi = \Delta\varphi(1 - r^2/r_0^2)$ and $f(\varepsilon) \sim 1/(\varepsilon + e\varphi_i)$, (where $\varphi_i$ is the potential of gas ionization, $\varphi < \varphi_i$), the equation (5) yields

$$\Delta\varphi = \sqrt{3L} \left(\frac{M}{m}\right)^{\frac{1}{2}} \left(\frac{1}{n_0 \sigma_i}\right)^{\frac{1}{2}} \left(\frac{\varphi_i}{U_0}\right)^{\frac{1}{2}} e n_-^{\frac{1}{2}} \left(1 - \frac{n_{0cr}}{n_0}\right)^{\frac{1}{2}}, \qquad (7)$$

Here, the voltage $U_0$ determines the energy $eU_0$ of the beam ions, and $M$ is the ion mass. Given the parameters of both the beam and the gas, one can calculate the $\Delta\varphi$.

It is to be noticed that the heated electrons do not carry out any energy from the beam when overcoming the barrier $e\Delta\varphi$ because their kinetic energy transforms to the electric field energy. This accumulated energy is converted to the kinetic energy of positive ions, which follow the electrons to establish a stationary state of the beam. Finally, the positive ions produced by gas ionization are particles that transport energy from the beam.

The authors of Ref. [4] have a different opinion on the physical mechanisms that are responsible for the formation of a steady state. With the objective of determining these mechanisms, we have numerically investigated the process of establishing the steady-state taking into account the Coulomb collisions both between beam ions and plasma electrons and between beam ions and positive plasma ions. The ion-electron collisions prove to be a major factor at high pressure, $n_0 > n_{0cr}$, and the ion-ion collisions are that at low pressure, $n_0 < n_{0cr}$. The results obtained were found to be in good agreement with both analytical estimates and experimental data.

Our numerical simulation was based on the particle-in-cell (PIC) method. A beam of negative $H^-$ ions is assumed to pass along the axis of a cylindrical metal tube whose radius is $R = 7.5\ cm$. The tube contains the neutral argon gas at the pressure $p = 10^{-5} - 1.3 \times 10^{-3}\ Torr$. The radius of the beam, $r_b$, is supposed to be constant, $r_b = 2.5\ cm$, and the ion energy is $15\ keV$. The bulk generation rate for electrons is $\gamma_e = n_0 n_- v_- (\sigma^{(i)} + \sigma^{(n)})$, and that for positive ions is $\gamma_i = n_0 n_- v_- \sigma^{(i)}$ ($\sigma^{(i)} = 3.5 \times 10^{-16}\ cm^{-2}$ is the ionization cross-section of the neutral gas, $\sigma^{(n)} = 2.5 \times 10^{-15}\ cm^2$ is the neutralization cross-section of the beam ions). The value of the stationary voltage drop in the beam depends on the speed distribution of the generated particles. The best agreement with the experimental data is obtained when the electron distribution function is assumed to be Maxwellian, and the energy of the positive ions is vanishingly small. Because the electrons and the positive ions are being heated by the beam ions, the linear momenta of these particles are changed in our calculation scheme at each time step according to the powers of heating. The boundary conditions for electric potential obtained by integrating Poisson's equation are $\left.\frac{\partial \varphi}{\partial r}\right|_{r=0} = 0,\ \varphi(r = R) = 0$.

Let us initially focus on the results of calculations at high pressures at first when the potential in the beam is positive. Fig.1 demonstrates the dependencies of the potential at the beam center (curve 1) and the radial voltage drop, $\Delta\varphi = \varphi(0) - \varphi(r_0)$, (curve 2) on gas pressure. The critical pressure at which the potential $\varphi(0) = 0$ is $p_{cr} = 1.65 \times 10^{-5}\ Torr$. The estimation (3) gives $p_{cr} \approx 1.8 \times 10^{-5}\ Torr$. As the pressure increases, the stationary-state time, $\tau_s$, decreases to $\tau_s \approx 5 \times 10^{-7}\ s$ at $p \approx 1.3 \times 10^{-3}\ Torr$. In case the Coulomb interaction is neglected in the numerical scheme, the value $\Delta\varphi = 0$ at any pressure.

Fig. 2 presents the potential distribution, $\varphi(r)$, at the pressure $p = 1.3 \times 10^{-3}\ Torr$. The corresponding stationary electron energy distribution (See Fig.3) proves to be close to Maxwellian at energies lower than the well depth. For energies higher than the well depth it abruptly drops to zero. So, the numerically obtained energy distribution is a distribution that serves as a basis for the approximate energy balance (6). Finally, Fig.4 shows the dependencies of the potential at the beam center (curve 1) and the voltage drop (curve 2) on the beam current density, $j_-$. Both of these dependencies are close to $\sim \sqrt{j_-} \sim n_-^{\frac{1}{2}}$, as expected from the estimate (7).

Now let us consider results of calculations at pressures lower than critical one. In this case the equilibrium potential in the beam is negative, and its minimum value is determined by Coulomb interaction between the negative ions of the beam and the positive ions of the plasma. Due to large mass of the positive ions, Coulomb heating occurs with small efficiency, and thus values of negative potential are small. Particularly, at $p = 10^{-5}\ Torr$ the numerically obtained voltage drop is only $\Delta\varphi \approx -0.043\ V$. As the pressure decreases, the stationary-state time increases to $80\ ms$ at $p = 10^{-5}\ Torr$.

A schema of the experimental setup for research on the spatial charge compensation is shown in Fig.5. The ion source is based on the reflective discharge. After the extractor *2*, the $H^-$-ions enter the magnetic field *3* whose strength is $2\ kO$. Finally, the magnetic lens *5* forms the beam *6* which propagates through the metal chamber to the target *10*. The chamber diameter is ~ 40 cm. The beam current in the chamber does not exceed $6 mA$, and the ion energies does not exceed $20 keV$. Fig.6 presents the value $\varphi(0)$ as functions of the gas pressure obtained for different gases filling the chamber. As expected, the potential $\varphi(0)$ is negative at low pressures, and positive at high pressures. The critical pressure is $p_{cr} = 3 \times 10^{-4}$, $4 \times 10^{-5}$, and $2 \times 10^{-5} Torr$ for helium, air, and krypton, respectively. The dependence of the critical pressure on the type of neutral gas is in qualitative agreement with the estimate (3). The measured values of the potential $\varphi(0)$ are close to the numerical results at high pressures. However, there is considerable discrepancy between these data at low pressures. The reasons why the experimental values are essentially lower than the calculated ones are discussed in the next section.

The points in Fig. 7 demonstrate the experimentally obtained dependence of the voltage drop $\Delta\varphi$ on the beam current density at high pressures. The solid lines present the dependencies that result from the estimate (7) for air and xenon. Taking into consideration the good agreement among the experimental (see Fig. 6 and 7), numerical (Fig.1-4), and approximate data (estimates (3) and (7)) we unambiguously conclude that:

1. It is the Coulomb collisions of the beam ions with the plasma electrons that define the degree of the overcompensation of the spatial charge in negative ion beams at high gas pressure.
2. The qualitative estimates of the overcompensation based on the energy balance equation for the trapped electrons (see Ref. [1, 3]), yield results that are close to the experimental and numerical data.
3. The results reported in [4] are incorrect because of an erroneously written energy balance.

## II. Dynamic decompensation of the beam due to current modulation by nonresonant oscillations

As noted in the previous section, the experimentally measured potential $\varphi(r = 0)$ is about 2 orders of magnitude higher than that obtained in our numerical model for pressures $p < p_{0cr}$. *Dynamic effects* can be naturally assumed to play a more significant role in this case.

The simplest of these is a decompensation due to nonresonant oscillations of the current caused by instabilities of the ion source plasma. Let $I = I_-(1 + \alpha \cos \omega t)$ be the beam current and let the amplitude $\alpha$ of the current oscillations be constant in time. If the frequency $\omega$ is high enough, so that the quantity of compensating particles (positive ions) created in the beam per half-cycle, $\pi/\omega$, per $1\ cm^3$ is essentially less than amplitude of oscillation of the beam density,

$$n_-^{(0)} v_- n_0 \sigma_i \frac{\pi}{\omega} \ll \alpha n_-^{(0)}, \qquad (8)$$

(where $n_-^{(0)}$ is the average ion beam density)

then the alternating beam spatial charge proves to be totally non-compensated, and the alternating potential $\tilde{\varphi}$ of the beam is

$$\tilde{\varphi} \approx \alpha \frac{I_-}{v_-} \qquad (9)$$

This phenomenon, which is conventionally called 'dynamic decompensation', can explain the abnormal potential $\varphi(0)$ at low pressures, when compared to the numerical results. Really, the frequency $f = \omega/(2\pi)$ and the amplitude $\alpha$ are correspondingly $200-500 kHz$ and $0.1-0.2$ in our experiments. The inequality (8) is valid for these parameters, and the alternating potential (9) was expected to be observed. Measurements of amplitude of the alternating potential by means of special capacitive probes have shown that there exist oscillations in the system, the amplitudes of which are close to the potentials measured by the incandescent probe in Ref. [1,3]. Therefore, the dynamic decompensation for our experiments is certain to be determined by the nonresonant oscillations of the beam current.

### III. Collective processes in the intense beams of negative ions

Collective oscillations in intense negative ion beams have a number of peculiarities in comparison with both electron beams and positive ion beams. However, as observed in both of these beam systems, two oscillation branches are excited in the system under study as well: electron (high frequency) and ion (low frequency) oscillations.

Unlike the case of current oscillations considered in the previous section, the collective oscillations are excited either in the beam itself, or (if perturbations at respective frequencies occur in the ion source plasma) these intensify in the drift space of the beam. In both cases, the oscillation amplitudes grow exponentially in the direction of the beam propagation (in the space surrounding the ion beam) during the initial stage of the instability development, which is described by linear equations. As the amplitude increases, nonlinear effects limit the growth. Steady-state oscillations with large amplitudes can essentially modify the following beam transport properties: the angular divergence, the degree of space charge compensation, and the beam emittance.

One of the authors of the present review studied in detail both the linear and the nonlinear stages of ion and electron oscillations in the plasma formed by the negative ion beam. The main characteristics of the linear oscillations (frequencies and space increments) and the main nonlinear mechanisms that limit the instability amplitudes were determined. A full summary of these researches is given in [1].

The most important practical conclusion, which results from above mentioned researches, is that the electron oscillations in the plasma for a wide range of pressures, and ion oscillations in the plasma at pressures greater than the critical pressure, do not lead to serious beam transport problems. These cause just a relatively small increase in the emittance and the degree of decompensation, which results in a minor increase of the beam angular divergence.

A different situation occurs for gas pressures lower than $p_{0cr}$, when the system is practically two-component and consists of negative ions of the beam and positive ions of the plasma that partially compensate the beam space charge. Electrons created by the ionization are repelled by the electric field; hence, their concentration is lower by several orders of magnitude than the ion concentration. The role of electrons in the development of ion oscillations in the plasma is known to be extremely important. These electrons shield the charge of all long-wave oscillations and enable intensification of those oscillations whose wavelengths are less than the Debye radius, $d_e$, of the electrons. Because the Debye radius usually exceeds the beam radius at low pressures, electrons cannot shield even the most long-wave perturbations (in radially-limited systems the longest wavelength does not exceed the doubled beam diameter). This circumstance, $d_e > r_0$, is not principal for the linear stage of the instability when its increment is weakly dependent on the wavelength. It does become important for the nonlinear stage because of the significant charge separation in the ion waves. Actually, estimates of the nonlinear effects indicate that the maximum amplitude of the electric field is proportional to total space charge per wavelength; consequently, the maximum amplitude of the potential in the wave is proportional to the wavelength squared. Therefore, ratio between the maximum amplitude in the system at low pressures (when electrons are absent) and that at high pressures (when electrons are present) is $(r_0/d_e)^2$.

The peculiarities of the ion oscillations observed at low pressures have been studied in experiments with comparatively intense ion beams ($I_- = 40\,mA$, Ref. [5]). Fig.8 presents oscillograms of the unstable voltage oscillations observed in [5] for different gas pressures. One can see from the figure that for $n_0 < n_{0cr}$ the voltage amplitude reaches $90V$ that comprises 30% of the beam potential calculated assuming the absence of the space charge compensation. This amplitude does not depend on the pressure up to $n_0 \sim n_{0cr}$. For $n_0 > n_{0cr}$ oscillations abruptly drops in accordance with the estimates.

A convincing demonstration that the ion oscillations mainly determine the potential distribution for $n_0 < n_{0cr}$ is given in Fig.9. This figure presents the flow of the positive ions to the chamber sidewall as function of the pressure $p$. The flow was measured at two different distances, $Z$, from the beam entrance to the chamber (See the curves 1 and 2). One can see that, in the region of the small amplitudes of the voltage oscillations, $n_0 > n_{0cr}$, these flows are about equal because they are only determined by the ionization rate in the measurement area. In contrast, in region II, $n_0 < n_{0cr}$, the flow at the shorter distance (curve 1) is considerably less than that at the larger distance (curve 2). Undoubtedly, the reason is that the ion oscillation amplitude sharply increases along the beam at low pressures to intensify the radial ion flow.

The experimental results presented show that at low gas pressures ion oscillations may have catastrophic influence on the beam transport. The extent of the influence grows as the beam current increases. One can easily see that the beam of $H^-$ ions whose current is about 100 $mA$ no longer can be transported when $n_0 < n_{0cr}$ due to the strong charge separation in the ion waves.

Thus, for intense beam transport, the gas pressure has to be higher than the critical value, $n_0 > n_{0cr} = 2v_i /(v\_r_0 \sigma_i)$. However, this criterion is sometimes hard to achieve because of the ion loss at high gas pressures due to the intense charge exchange. In this case, one can use the suppression of the ion-ion instability proposed in Ref. [6]. The idea of this instability suppression is that the electrons are forcedly injected into the beam from a thermo emitter in order to shield the electric field of the nascent ion-ion waves.

Fig.10 shows the scheme of the setup for this suppression. A beam of $H^-$-ions with current $\sim 50$ $mA$ and energy $\sim 10$ $keV$ is extracted from the surface-plasma source, *1*. Magnetic field of up to $2kO$ created by the permanent magnets *2* bends and forms the beam. In the straight section, the beam propagates inside the two solenoids with currents of opposite directions. These solenoids form the cusp magnetic field to hold electrons in the beam. The beam collector *6* is a tantalum plate with holes each having a 0.3 mm diameter and placed with 5 mm spacing in the corners of squares. Images of the beams passing through the holes and falling on the luminescent screen *7* are used to measure the beam emittance. The electron emitter *9* is a tungsten thread heated by an electric current. The voltage between the emitter and the chamber, and the emission current are about $\sim 150V$ and $\sim 250 mA$ correspondingly. The neutral gas in the chamber is argon.

The oscillograms of the column *1* in Fig. 11 demonstrate the potential oscillations for different gas pressures when the instability develops in the ion-ion plasma. After the emitter is turned on, the instability is suppressed by other equal conditions (See the oscillograms of the column *2* as contrasted to the column *1*). One can see that injection of electrons for $p < p_{0cr} \approx 5 \times 10^{-5}$ $Torr$ results in reducing the oscillation amplitudes to values observed at $p > p_{0cr}$.

Suppression of the instability leads to a significant voltage decrease (that is, decompensation degree) that decreases the emittance and the angular divergence of the beam.

**Conclusions**

The results of our research at the Institute of Physics of National Academy of Science of Ukraine are as follows:

1. Approximate analytical expressions based on the energy balance for the plasma electrons and ions have been obtained to evaluate the minimal electric field in the compensated negative ion beams for different gas pressures.

2. Numerical calculations of the steady state of the beams have been performed using a PIC-code. The Coulomb collisions of the negative ions with the plasma electrons (at high pressures) and the plasma ions (at low pressures) were shown to determine the physical mechanisms forming the equilibrium state.

3. Experiments using an $H^-$-ion beam reveal that the radial electric fields measured at high gas pressures are close to both the numerical data and the estimated values. However, at low gas pressures, instabilities in the ion source and the beam plasma can cause the dynamical beam decompensation which increases the extent of the decompensation as compared with our numerical results.

4. The ion-ion oscillation can have catastrophic influence on the beam transport at low pressures. One method of suppression of this instability is proposed and demonstrated in experiments with the intense beams of $H^-$ ions. The idea of this method is that the electrons forcedly injected to the beam plasma shield the electric field of the nascent ion-ion waves to prevent their growth.

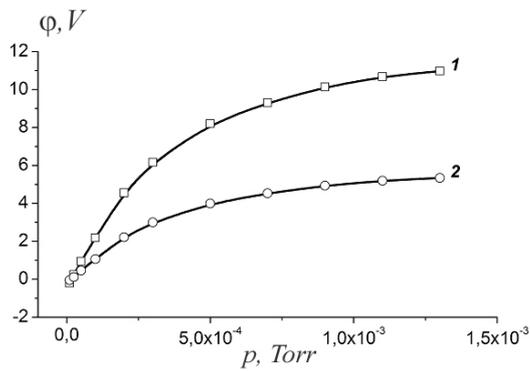

Fig.1. Pressure dependencies of the potential at the beam center (curve 1) and the radial voltage drop in the beam (curve 2).

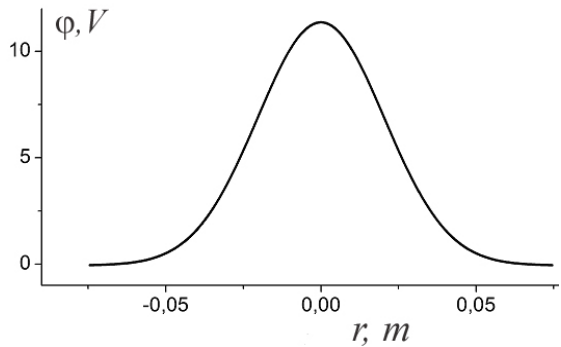

Fig.2. The radial potential distribution at pressure $p = 1.3 \times 10^{-3} Torr$.

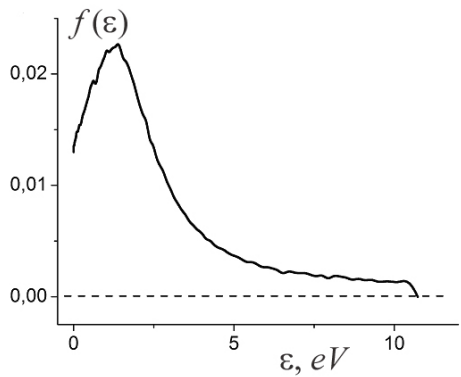

Fig.3. The steady-state energy distribution function of electrons at $p = 1.3 \times 10^{-3} Torr$.

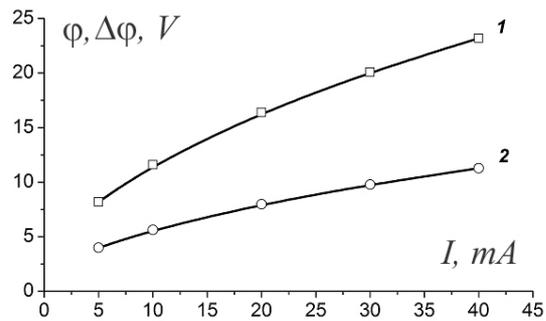

Fig.4. Dependencies of the potential at the beam center (curve 1) and the voltage drop in the beam (curve 2) on the beam current density.

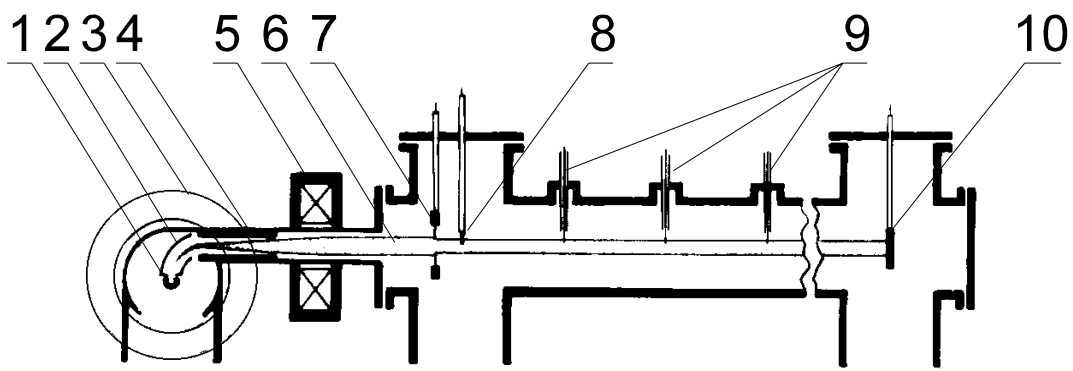

Fig.5. Scheme of experimental setup. 1 – ion source based on reflective discharge; 2 - extractor; 3 – magnetic field; 4 – magnetic shield; 5 – magnetic lens; 6 – ion beam; 7 - aperture; 8,9 - probes; 10 - target.

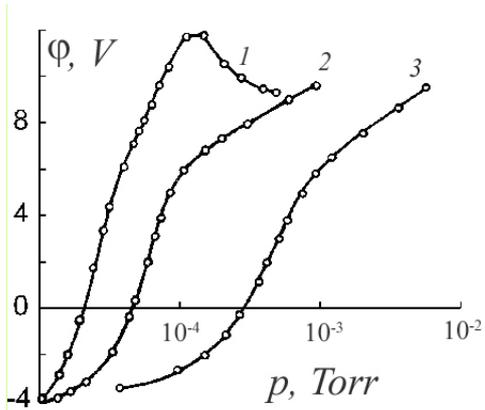
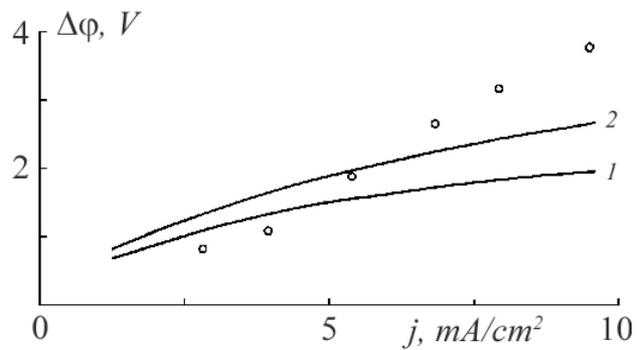

Fig.6. Experimental dependencies of the potential at the beam axis on gas pressure. 1 - krypton, 2 – air, 3 – helium; $I_- = 4\ mA,\ eU_0 = 10\ keV$.

Fig.7. The radial voltage drop as function of the beam current density; curves 1 and 2 are calculated for the air and the krypton correspondingly; the points are the experimental data for krypton.

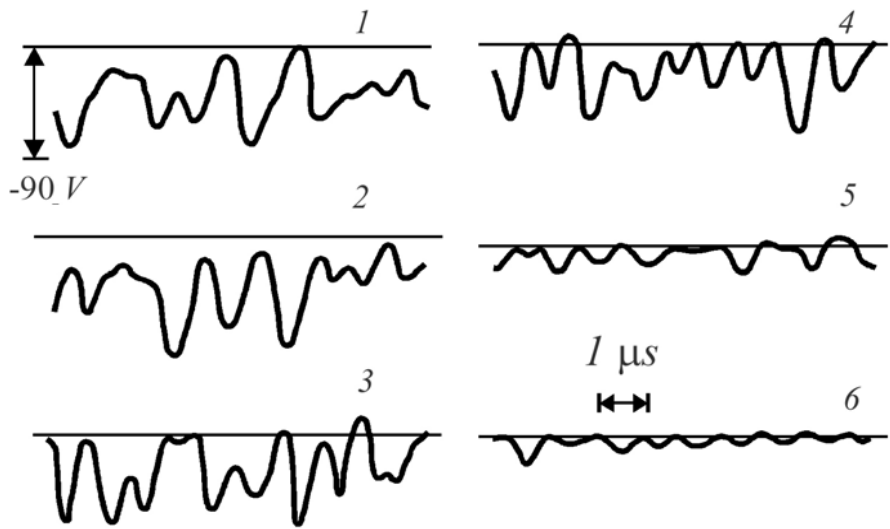

Fig.8. Oscillograms of unstable ion oscillations of the plasma; $I_- = 40\ mA,\ eU_0 = 14\ keV$. For the curves 1–6 the pressure is $5\cdot10^{-6}$, $1.4\cdot10^{-5}$, $2.8\cdot10^{-5}$, $4\cdot10^{-5}$, $6\cdot10^{-5}$, and $6.8\cdot10^{-5}$ Torr correspondingly.

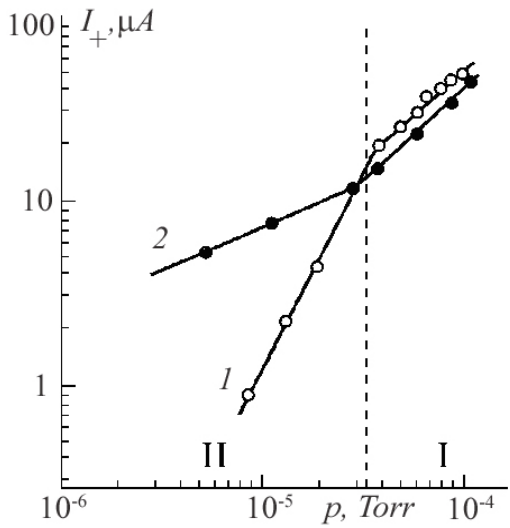

Fig.9. Pressure dependencies of the radial current of the positive ions on pressure for two cross sections of the beam: $Z = 18$ cm (curve 1) and $Z = 38$ cm (curve 2); $I_- = 40\ mA,\ eU_0 = 14\ keV$

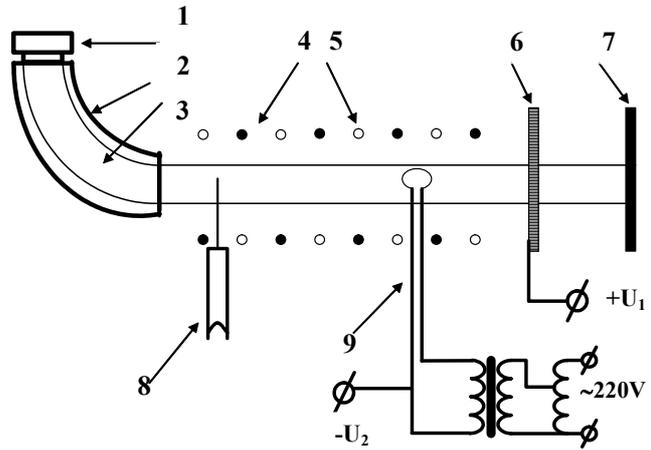

Fig.10. Experimental setup for stabilization of the ion-ion instability. 1 - source of H⁻ ions; 2 – beam turning magnets; 3 - H⁻ ion beam; 4,5 - turns of solenoid coils; 6 - collector; 7 - luminescent screen; 8 - capacitive probe; 9 - electron emitter circuit.

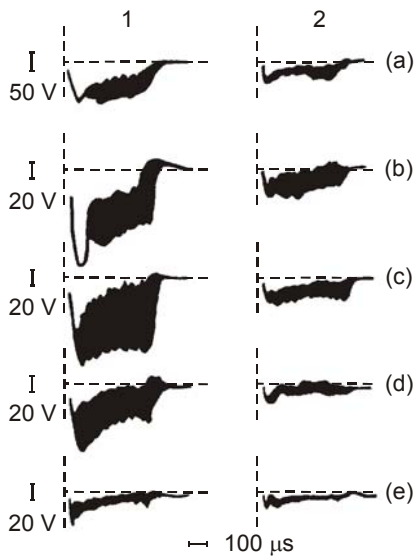

Fig.11. Oscillograms of ion oscillations of the plasma potential for different pressures without stabilization (column 1) and with the stabilization (column 2). The solenoid current is $200\ A$; the electron emission current is $250\ mA$; gas pressure $p = 5\cdot10^{-6}\ Torr$ (a); $9\cdot10^{-6}$ (b); $2.4\cdot10^{-5}$ (c); $3.9\cdot10^{-5}$ (d); $7.7\cdot10^{-5}$ (e).